\documentclass[reprint,superscriptaddress]{revtex4-2}
\usepackage[T1]{fontenc}
\usepackage{siunitx}
\newcommand{\abs}[1]{\ensuremath{\lvert#1\rvert}}

\usepackage{graphicx}
\usepackage{amsmath}
\usepackage{amssymb}
\usepackage{multirow}

\newcommand{\affilqmt}{\affiliation{Quantum Motion, 9 Sterling Way, London, N7 9HJ, United Kingdom}}
\newcommand{\affilucl}{\affiliation{Department of Electronic and Electrical Engineering, UCL, London, WC1E 6BT, United Kingdom}}
\newcommand{\affillcn}{\affiliation{London Centre for Nanotechnology, UCL, 17-19 Gordon Street, London, WC1H 0AH, United Kingdom}}
\begin{document}

\title{Rapid cryogenic characterisation of 1024 integrated silicon quantum dots}

\author{Edward J.~Thomas}
\thanks{These authors contributed equally}
\affilqmt
\affilucl

\author{Virginia N.~Ciriano-Tejel}
\thanks{These authors contributed equally}
\affilqmt

\author{David F.~Wise}
\affilqmt

\author{Domenic Prete}
\affilqmt

\author{Mathieu de Kruijf}
\affilqmt
\affillcn

\author{David J.~Ibberson}
\affilqmt

\author{Grayson M.~Noah}
\affilqmt

\author{Alberto Gomez-Saiz}
\affilqmt

\author{M.~Fernando Gonzalez-Zalba}
\affilqmt

\author{Mark A.~I.~Johnson}
\email{mark@quantummotion.tech}
\affilqmt

\author{John J.~L.~Morton}
\affilqmt
\affilucl
\affillcn

\begin{abstract}
Quantum computers are nearing the thousand qubit mark, with the current focus on scaling to improve computational performance~\cite{arute_quantum_2019, Zhong2019, Madsen2022, Acharya2023, Kim2023}. 
As quantum processors grow in complexity, new challenges arise such as the management of device variability and the interface with supporting electronics~\cite{gonzalezzalba2021, Zhang2022}. 
Spin qubits in silicon quantum dots are poised to address these challenges with their proven control fidelities~\cite{Xue2022, Noiri2022, Mills2022} and potential for compatibility with large-scale integration~\cite{xue_cmosbased_2021,ruffino_integrated_2021}. 
Here, we demonstrate the integration of 1024 silicon quantum dots with on-chip digital and analogue electronics, all operating below \qty{1}{\kelvin}.
A high-frequency analogue multiplexer provides fast access to all devices with minimal electrical connections, enabling characteristic data across the quantum dot array to be acquired in just 5 minutes.
We achieve this by leveraging radio-frequency reflectometry with state-of-the-art signal integrity, reaching a minimum integration time of \qty{160}{\pico\second}.
Key quantum dot parameters are extracted by fast automated machine learning routines to assess quantum dot yield and understand the impact of device design.
We find correlations between quantum dot parameters and room temperature transistor behaviour that may be used as a proxy for in-line process monitoring. 
Our results show how rapid large-scale studies of silicon quantum devices can be performed at lower temperatures and measurement rates orders of magnitude faster than current probing techniques~\cite{neyens2023probing}, and form a platform for the future on-chip addressing of large scale qubit arrays.
\end{abstract}

\maketitle

\begin{figure*}[t]%
\centering
\includegraphics[width=\textwidth]{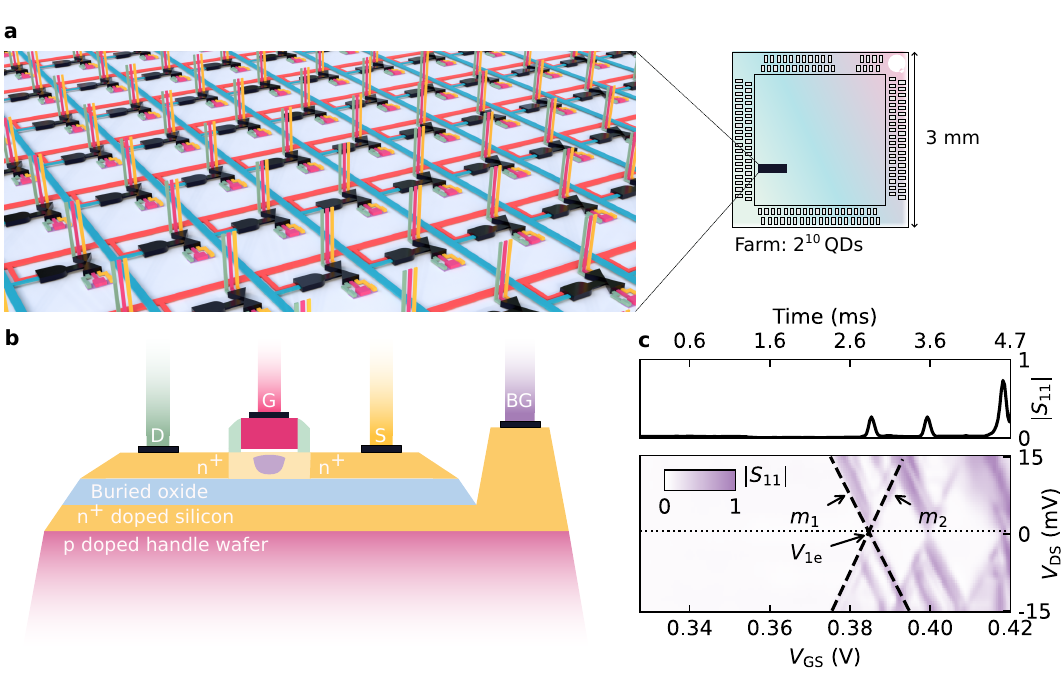}
\caption{
\textbf{a}, 3D schematic render of the 1:1024 multiplexer, with analogue access (green, pink, yellow) to each quantum dot device controlled by row-column addressing (red and blue wires). 
This farm of devices occupies a small section of a $\qty{3}{\milli\metre} \times \qty{3}{\milli\metre}$ silicon die.
\textbf{b}, Schematic cross-section of a single transistor with a quantum dot (purple) below the gate and situated between the drain and source.
The region where the quantum dot forms is undoped silicon.
\textbf{c}, Example 2D map showing a normalised device response (colour scale) as a function of source-drain and gate voltages. The dashed line shows an automated fit to the first measured Coulomb blockade oscillation.
The top panel shows a line cut at $V_{\rm DS} = \qty{0}{\volt}$ (indicated by the dotted line), with the time-axis aligning with the voltage axis in the bottom panel.
}
\label{fig1}
\end{figure*}

\section*{Introduction}

Semiconductor quantum dots (QD) are a promising platform to implement a fault-tolerant quantum computer, a novel computing paradigm expected to greatly outperform classical high-performance computers in areas such as materials and drug discovery, optimisation and machine learning~\cite{montanaro_quantum_2016}.
This is afforded by their small footprint, ability to host highly-coherent and controllable spin qubits, and their potential compatibility with advanced semiconductor manufacturing.
In particular, spin qubits in isotopically enriched silicon have demonstrated control, preparation and readout fidelities~\cite{Mills2022,yoneda2020quantum,blumoff2022fast} above the threshold to perform quantum error correction (QEC)~\cite{fowler2012surface}.
This long term goal of fault-tolerance, where an error-correcting code reduces the error rate to a negligible amount, is predicted to require millions of physical qubits to solve practical problems~\cite{beverland_assessing_2022}.

As solid state quantum processors scale up to useful levels of complexity, two important challenges must be addressed. 
First, the number of connections between room temperature and the quantum processor cannot continue to grow in proportion with the number of qubits~\cite{franke_rent_2018, vandersypen_interfacing_2017,gonzalez-zalba_probing_2015}. 
Frequency-division multiplexing has been applied to allow multiple qubits to share measurement electronics, however, frequency crowding limits this approach -- so far to 8 qubits per line~\cite{George2017}.
Crossbar approaches~\cite{bavdaz_quantum_2022,li_crossbar_2018}, in which $O(\sqrt{N})$ lines intersect at $N$ qubit locations, offer an elegant solution to reducing wiring, albeit with stringent requirements on qubit variability and limitations in the operation of the processor. 
Ultimately, the use of switches to achieve time-division multiple access (TDMA), as in dynamic random access memory (DRAM), in both the read-out and control lines to each qubit unit cell provides the greatest flexibility and scalability. 
For dc signals, off-chip and on-chip cryogenic switches with ratios up to 1:36 and 1:64 respectively have been used to address quantum device arrays~\cite{bavdaz_quantum_2022,al-taie_cryogenic_2013,puddy_multiplexed_2015,lee_16_2022}. Multiplexed control circuitry operating at \SI{1}{\K} with $>\SI{100}{\MHz}$ pulsing has been recently reported~\cite{jadot_cryogenic_2023}. For radio-frequency (rf) signals used for high-speed read-out, up to 1:3 switching has been shown on-chip~\cite{ward_integration_2013,schaal_cmos_2019,ruffino_integrated_2021} while high-frequency 1:4 cryogenic multiplexers have been used with superconducting qubits~\cite{acharya_multiplexed_2023}.
Developing TDMA at scale requires an efficient interface between classical electronics and the quantum processor.

A second challenge is managing and minimising process variability between the qubits \cite{gonzalezzalba2021}, requiring each qubit to be independently characterised and tuned.
Minimising process variability is already an integral component of modern semiconductor manufacturing, where it has been tamed even at the extreme nanometre length scales, where silicon device fabrication is pushed to its limit. 
While current process characterisation is optimised for classical transistors, extending this to quantum devices requires substantial development in high throughput cryogenic testing capabilities. 
State-of-the-art methods rely on newly developed cryogenic probing to enable wafer-scale testing~\cite{neyens2023probing}. 
However, this method is currently limited to temperatures above \qty{1.6}{\kelvin} and is unable to utilise all wafer space since devices need to be directly contacted to macroscopic pads.
On-chip multiplexing techniques can provide access to large numbers of densely packed devices with minimal input/output connections, while testing can be performed at the optimal temperature conditions of spin qubit devices (mK)~\cite{ruffino_integrated_2021,bavdaz_quantum_2022}. 
Large scale on-chip switching of rf signals therefore addresses both short-term and long-term challenges in quantum computer development: providing a way to characterise many quantum devices to address process variability, and to address many qubits in scaled up quantum processors. 

In this work, we rapidly characterise a farm of 1024 QD devices, fabricated using a commercial foundry process by integrating high-speed classical multiplexing electronics to individually address each device.
We develop and deploy tools to automatically extract key indicators for QD performance and their suitability for use in qubit technologies. 
We perform a statistical analysis of different device dimensions, and under varied operating conditions. 
Ultimately, we show that all of these devices can be well characterised at cryogenic temperatures with less than 5 minutes of measurement time by means of radio-frequency reflectometry techniques at the state-of-the-art.
Finally, we establish a link between cryogenic and room-temperature device properties, opening a new avenue for pre-cryogenic validation of silicon qubit technologies.

\section*{Cryogenic multiplexed access to 1024 quantum dots}

We use an analogue bus of three control lines and 10 digital address lines (5 row select and 5 column select lines) to address each QD device-under-test in a~\numproduct{32x32} array via a 1:1024 multiplexer.
The devices are selectively connected to the analogue bus using CMOS transmission gates integrated within the same silicon as the quantum devices. 

The IC is designed using an ultra-thin body (UTB) and buried oxide (BOX) fully-depleted silicon-on-insulator (FD-SOI) process. 
Quantum dots form in the undoped channel of the transistors when a voltage difference $V_{\rm GS}$ between the gate (G), and source (S) approaches the threshold voltage, see Fig.~\ref{fig1}b.
To observe discrete charging of the QDs, a source and drain (D) tunnelling resistance larger than the resistance quantum is necessary. 
This occurs naturally in these devices due to the increase in resistivity of the undoped silicon region at low temperatures.
This region lies beneath the gate and spacers (Fig.~\ref{fig1}b).
In Fig.~\ref{fig1}c, we see an example of the characteristic discrete charging, i.e.~diamond shaped regions of decreased conductance nestled within regions of higher conductance.


\section*{Reflectometry performance via a multiplexer}

\begin{figure}[t]%
\centering
\includegraphics[width=\linewidth]{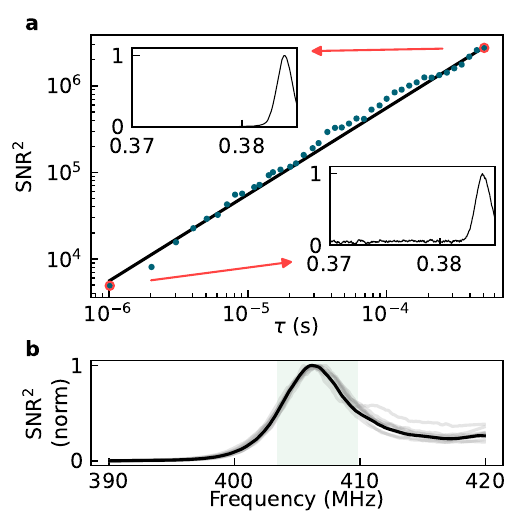}
\caption{
    \textbf{a}, Power signal-to-noise ratio as a function of integration time. 
    The insets show example Coulomb oscillations, with normalised amplitudes, measured at the maximum and minimum integration times (indicated in red).
    The inset horizontal axes are the gate voltage, $V_{\rm GS}$ in Volts.
    A linear fit (solid black line) gives a minimum integration time $t_{\rm min} \approx \qty{0.16}{\ns}$. 
    \textbf{b}, The normalised signal-to-noise ratio for nine devices are each shown as faint grey lines, and the heavy black line is the mean response. The green shaded region is the mean bandwidth over the selected devices. 
}
\label{fig2}
\end{figure}

\begin{figure}
    \centering
    \includegraphics[width=\linewidth]{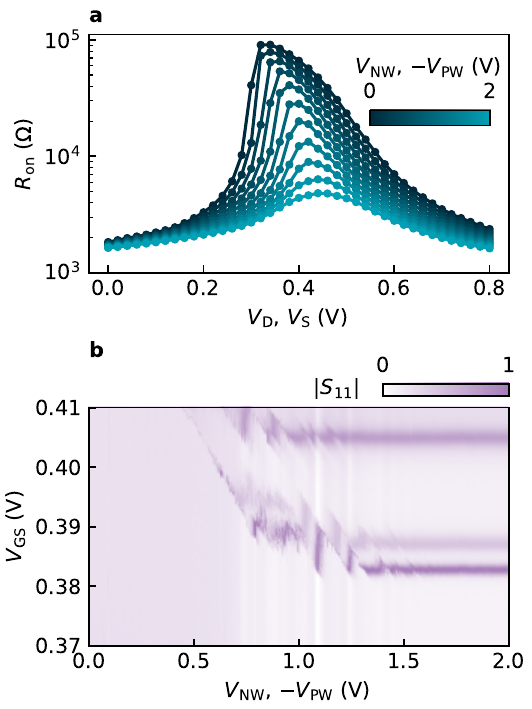}
    \caption{\textbf{a}, The on-resistance $R_{\rm on}$ of a single multiplexer transmission gate with varied analogue back-gates $V_{\rm NW}$ and $V_{\rm PW}$ for the NFET and PFET respectively.
    \textbf{b}, The normalised reflected signal from the device as the analogue back-gates are varied. The multiplexer reaches a stable configuration when $V_{\rm NW} > \qty{1.5}{\volt}$ and $V_{\rm PW}< \qty{-1.5}{\volt}$, corresponding to a transmission gate impedance of $\approx\qty{2}{\kilo\ohm}$ with a $V_{\rm DS} = \qty{20}{\milli\volt}.$
    }
    \label{fig:mux}
\end{figure}

To expedite device measurements, we use radio-frequency (rf) reflectometry~\cite{schoelkopf_radiofrequency_1998, duty_observation_2005, wallraff_strong_2004a, vigneau_probing_2023}. 
Reflectometry can detect quantum dot charge transitions with high bandwidth by measuring changes in the device impedance.
The typically high device impedance (\SI{>100}{\kilo\ohm}) is matched to a \qty{50}{\ohm} line by embedding the IC in a matching network containing a superconducting spiral inductor (see Methods). 
This allows us to monitor the reflected voltage as a function of the device impedance.

\begin{figure*}[t]%
    \centering
    \includegraphics{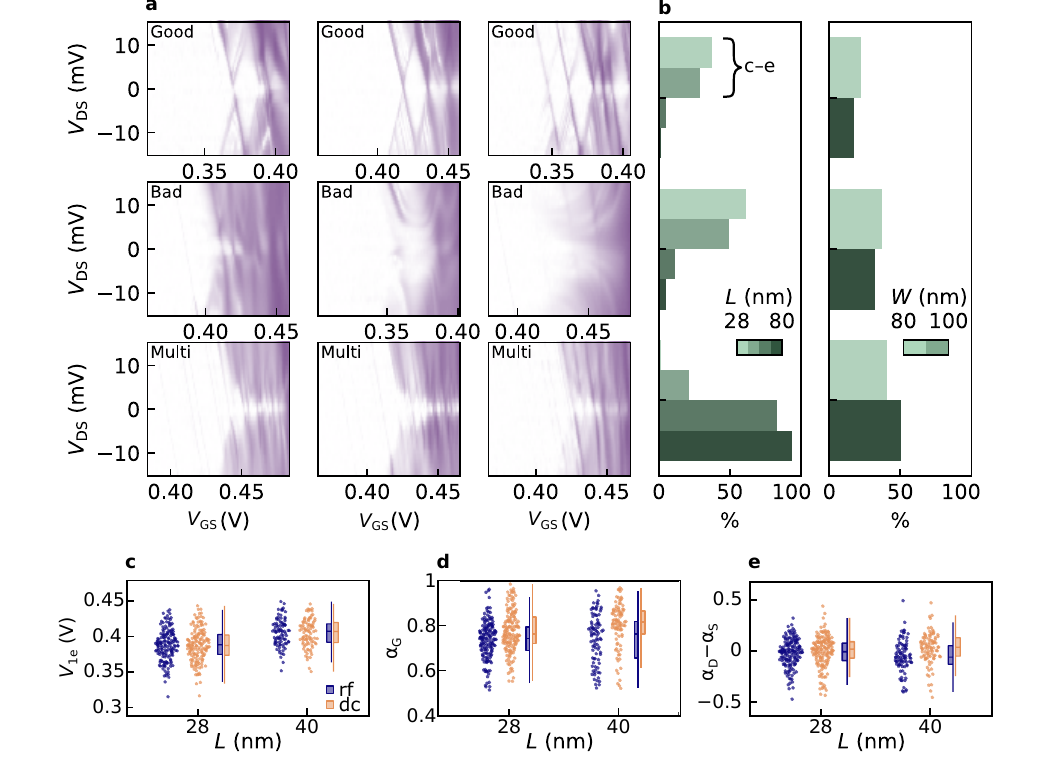}
    \caption{
        \textbf{a}, Example 2D maps for 9 different devices. Labels indicate the quality of Coulomb blockade observed; ``multi'' refers to signatures of multiple series dots forming in the device. 
        \textbf{b}, Relative frequency of each device category for different gate lengths $L$ and channel widths $W$. 
        \textbf{c-e}, Distributions of automatically extracted first-electron voltage (\textbf{c}), gate lever arm (\textbf{d}) and drain-source coupling asymmetry (\textbf{e}) for different gate lengths.
    }\label{fig3}
\end{figure*}

Here, we characterise the performance of this technique in terms of the signal-to-noise ratio (SNR) and bandwidth at a farm level. 
A peak in the $S_{11}$ trace corresponds to the electrochemical potential of the QD being in alignment with the Fermi level of the source or drain. 
We define the signal as the height of this peak, and the noise as the standard deviation of the background signal.
In Fig.~\ref{fig2}a, we show the SNR$^2$ of a QD charge transition as a function of integration time $\tau$. Through linear extrapolation,  we determine the minimum integration time required to attain an SNR of 1, which amounts to $t_{\rm min} = \qty{160}{\pico\second}$. 
This result demonstrates that the multiplexer does not compromise the signal quality, in fact, our apparatus outperforms the state-of-the-art achieved with reflectometry in single electron transistors~\cite{keith_singleshot_2019}.

Next, we test the bandwidth at a farm level by looking at the dependence of the SNR with respect to the probe frequency for 9 exemplary devices, see Fig.~\ref{fig2}b. 
For TDMA, it is critical that the frequency region of high signal overlaps for each device in the farm.
We demonstrate that this holds, with the average SNR bandwidth (\qty{6.4}{\mega\hertz}), defined as the full-width at half-maximum of the SNR$^2$ signal, which is similar to the resonator bandwidth (\qty{9.5}{\mega\hertz}).
Overall, the TDMA implementation presented here should comfortably result in QD measurements with a ${\rm SNR} > 20$ with an integration time per point of \qty{100}{\nano\second}. 
In this work the lowest integration time we use is $\qty{1}{\micro\second}$.

Finally, we show how the cryogenic performance of FD-SOI~\cite{beckers2019characterization,paz2020performance} is pivotal in creating our low temperature multiplexing circuit.
In particular, back-gating through the BOX enables compensation of the known transistor threshold voltage increase at low temperatures which becomes important when delivering high-frequency signals through the MUX (Fig.~\ref{fig:mux}). 
Notably, we measure the on-resistance $R_{\rm on}$ of a single transmission gate over a wide back-gate voltage ($V_{\rm NW,\,PW}$ for the NFET and PFET, respectively) range and for multiple common-mode drain/source voltages.
During these measurements the NFET gate is held at $V_{\rm DD} = \qty{0.8}{\volt}$ and the PFET gate is held at $V_{\rm SS} = \qty{0}{\volt}$.
By applying a positive back-bias, we can reduce the on-resistance by more than an order of magnitude for a common-mode voltage $V_{\rm D} = V_{\rm S} = \qty{0.4}{\volt}$.
The impact of the back-biasing is also evident from the quality of the reflectometry signal.
Fig.~\ref{fig:mux}b shows two Coulomb oscillations as measured in reflectometry while the multiplexer back-bias is increased.
The two oscillations are well resolved when $V_{\rm NW}=-V_{\rm PW} > \qty{1.5}{\volt}$. 
However, for lower values a substantial voltage drop occurs at the MUX, shifting the position of the oscillations. 
The shift is accompanied by a reduction of the SNR, which, along with the unstable signal behavior, can be linked to increasing MUX resistance.

\section*{Robust analysis and extraction of quantum dot features at scale} \label{sec:robust-analysis}
To characterise the QD devices, we extract (see Methods) the first observed electron loading voltage ($V_{\rm 1e}$), the gate lever arm ($\alpha_{\rm G}$) which describes the strength of the electrostatic coupling of the gate electrode to the dot, and the drain-source lever arm difference ($\alpha_{\rm D} - \alpha_{\rm S}$), to measure the device asymmetry. 
These parameters are the ones that can be extracted unambiguously from a single (the first) Coulomb blockade oscillation. 
We expect these parameters will vary from device to device as the shape and location of a dot may differ due to design differences or through process factors e.g.~surface roughness at the Si/SiO$_2$ interface~\cite{cifuentes2023bounds}, and trapped charges in the oxide or at the interface.
We define $V_{\rm 1e}$ as the gate voltage at which we first detect a Coulomb oscillation peak at zero $V_{\rm DS}$, representing the loading of an electron from the reservoir. 

To extract further electrostatic properties, like the charging energy, further oscillations are required, but given the disorder, the presence of additional dots cannot be conclusively ruled out. 
Furthermore, a remote charge sensor would be required to verify we have reached the single-electron regime, so here we instead find the distribution of gate voltages corresponding to the first transition observable using our charge sensing methods (in-situ reflectometry and dc transport), in order to establish the trends and variability between devices. 
Though we cannot guarantee reaching the single-electron regime, the Coulomb diamonds recorded are indicative of the few-electron regime. 

As stated above, a single measurement is used to extract all of these parameters, by monitoring the device as $V_{\rm GS}$ and $V_{\rm DS}$ are varied (see Methods).
A measurement of this kind is shown in Fig.~\ref{fig1}c.
In the farm, 8 transistor variants were tested with increasing gate lengths ($L$) \qtylist{28;40;60;80}{\nano\metre} and channel widths ($W$) \qtylist{80;100}{\nano\metre}.

Our first observation shows that not all transistors can provide good QDs (Fig~\ref{fig3}a).
For these devices, we therefore cannot extract the QD parameters described above, and so they must first be filtered out.
We have trained a convolutional neural network (CNN) to categorise our devices into three categories: clear Coulomb blockade (good), no Coulomb blockade (bad) and multiple series QDs (multi), which present as non-closing diamonds.
Each device is manually labelled by a domain expert and these labels are used to train the CNN (see Methods).
The proportion of devices that fall into each category is shown in Fig~\ref{fig3}b.

To automatically extract parameters we developed tools to process data acquired through both transport (dc) and reflectometry (rf) measurements.
Transistor characteristic measurements are commonly performed in dc; thus, the dc data presented here serves as a reference for comparison with the rf data, revealing a good agreement between both techniques (Fig.~\ref{fig3}c-e and Fig.~\ref{fig:back-gate}).
All parameters $V_{\rm 1e}$, $\alpha_{\rm G}$ and $\alpha_{\rm D} - \alpha_{\rm S}$ can be determined from the pair of intersecting lines in a charge map (see Fig.~\ref{fig1}c).
The intercept gives $V_{\rm 1e}$, and the lever arms can be calculated from the gradients of the two lines (see Methods).
For each charge map, we perform a fitting routine which finds the best pair of lines, with higher weight given to $V_{\rm 1e}$ at lower voltages, and intercept close to $V_{\rm DS} = 0$.

\section*{Variability of industrially fabricated quantum dots}
To assess inherent process variability, considerable effort was placed in the design stage to suppress known sources of semiconductor process variability, e.g.~layout effects (see Methods), so the variability we measure is primarily inherent to each DUT. 
As shown in Fig~\ref{fig3}b, in devices with greater gate lengths, multiple dots are generated, resulting in complex multi-dot structures. 
On the other hand, devices with shorter gate lengths yield single quantum dots, however in some devices the small gate length leads to an early transistor turn-on, resulting in `bad' dots.
Overall, the device designs with the highest proportion of good QD features have shorter gate lengths, \qtylist{28;40}{\nano\metre}. 
The two channel widths do not impart significant differences across the farm.
The extracted parameters for these good devices are shown in Fig.~\ref{fig3}c-e.
With decreasing gate length we see a lower threshold voltage due to the increasing effect of the electric field produced by the source and drain, a well-known short channel transistor effect, likely caused by drain-induced barrier lowering (DIBL)~\cite{troutman_simple_1977}.
The gate lever arm and lever arm asymmetry remain fairly constant, indicating that dots remain well controlled by the gate even at the smallest dimensions. 

For the $L = \qty{28}{\nano\metre}$ case, we find the first-electron voltages ($\overline{V_{\rm 1e}} = \qty{387\pm22}{\milli\volt}$) and the gate lever arms ($\overline{\alpha_{\rm G}} = \num{0.741(82)}$) are narrowly distributed.
We find the standard deviation of $V_{\rm 1e}$ ($\approx~\qty{22}{\milli\volt}$) is comparable to the spacing (\qty{25(4)}{\milli\volt}) between the first and second observed electron ($V_{\rm 2e}$) loading voltages, measured from a subset of devices where a second transition is clearly visible. 
This suggests that the tight requirements for shared voltage control~\cite{bavdaz_quantum_2022} are within reach, but need further reduction.
Alternatively, small variations may be compensated with independent voltage trimming of each of the QD back-gates, however this comes at the cost of a greater number of control lines per QD.
We note that the large gate lever arm will be beneficial when implementing gate-based dispersive readout\cite{vigneau_probing_2023}.

The mean dot asymmetry $\overline{\alpha_{\rm D} - \alpha_{\rm S}} = \num{-0.040(150)}$ show that on average the QDs are well-centered in the channel.
We note that, since all lever arms must add to 1, a large asymmetry places an upper bound on the gate lever arm, $\alpha_{\rm G} \le 1 - \abs{\alpha_{\rm D} - \alpha_{\rm S}}$ (see Supplementary).
This result highlights an inverse relation between asymmetric QD position within the channel and gate control over the QD, emphasising the importance of QD location being central under the gate.
We note that for the devices here, the quantum dot is placed between two large conducting leads, however for larger QD arrays most dots may only have a single lead, or only other dots nearby.
In such devices, the importance of the short-channel effect we see here is lessened.
In light of this, our measurements provide a worst-case indication for dot asymmetry under a single gate.

\section*{Room-temperature correlations with QD parameters}

\begin{figure}[t]%
\centering
\includegraphics{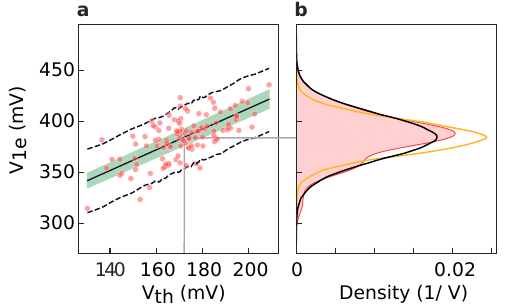}
\caption{ 
\textbf{a} $V_{\rm 1e}$, measured at $\qty{50}{\milli\kelvin}$, against $V_{\rm th}$, measured at room temperature. 
The solid black line is the average of the estimated linear fits and the shaded region shows the 95\% confidence interval over possible linear fits (slope and intercept). The dark dashed lines represent the 95\% confidence interval for $V_{\rm 1e}$ accounting for noise.
The solid grey line indicates a nominal $V_{\rm th}$ used to analyse the distribution of $V_{\rm 1e}$ in b.
\textbf{b} The extracted $V_{\rm 1e}$ distribution from measured data (red), shown against the predicted posterior distribution (black) of $V_{\rm 1e}$ including the observed $V_{\rm th}$ variation. 
The yellow line indicates a best-case distribution of $V_{\rm 1e}$ when $V_{\rm th}$ takes on a single value (centred at the $V_{\rm 1e}$ indicated by the grey line).
}
\label{fig:room-temperature-correlation}
\end{figure}

It would be ideal to be able to determine QD parameters without needing to cool the device to cryogenic temperatures. 
For a QD, once the thermal energy $k_{\rm B} T$ becomes much larger than the charging energy $E_C = {\rm e}^2 / C_\Sigma \sim \qty{15}{\milli\electronvolt}$ (see Supplementary), transport through the transistor does not exhibit blockade and the device behaves as a simple transistor.
This removes access to the first-electron voltage and the device lever arms, however we gain a new set of parameters used in classical modelling of transistors, such as the threshold voltage ($V_{\rm th}$). 

We next establish a direct link between QD parameters and classical transistor behaviour (see Supplementary) which allows device yield and uniformity to be assessed without requiring expensive and time-intensive cooling in a dilution refrigerator.
Recent work has correlated classical transistor behaviour from room temperature to \qty{4.2}{\kelvin}~\cite{Paz2020}, here we extend this to the behaviour of QDs at lower temperatures.
Furthermore, an increased subthreshold current at room temperature has been linked to the presence of quantum dots
~\cite{sellier_subthreshold_2007} or individual dopants~\cite{sellier_transport_2006,Wacquez_single_2010,deshpande_novel_2011} in the channel, observed in transport at cryogenic temperatures. 
We build on this work and highlight that pre-cryogenic validation can be an invaluable tool in the silicon quantum computer life cycle.

To understand the systematic relationship between $V_{\rm 1e}$ and $V_{\rm th}$ and their random variations, we employ probabilistic programming (see Methods).
With this approach we can disentangle the intrinsic variation of $V_{\rm 1e}$, i.e.~the distribution of $V_{\rm 1e}$ for a fixed $V_{\rm th}$, from the variation of $V_{\rm th}$ inherent to the CMOS fabrication process.
In principle, $V_{\rm th}$ variation could be reduced at a foundry level, allowing a corresponding reduction in $V_{\rm 1e}$ variation.
We stress that while these two variations may have the same physical origin, we do not model that here.
For clarity, here we use the term $V_{\rm th}$ as a single observed threshold voltage, and all presented data is collected from devices with the shortest gate length, $L = \qty{28}{\nano\metre}$.

We consider a simple linear model of the form
\begin{equation}
    \hat{V}_{\rm 1e} (V_{\rm th}) = \alpha V_{\rm th} + \beta + \mathcal{N}(0,~\sigma),
\end{equation}
where $\alpha$, $\beta$ and $\sigma$ are random variables, each chosen with a weakly informative normal $(\mathcal{N})$ prior distribution.
We use Hamiltonian Monte Carlo~\cite{Hoffman2011} to develop a posterior distribution that accurately matches the data shown in Fig.~\ref{fig:room-temperature-correlation}a.
The slope ($\overline{\alpha} =\num{1.01(2)}$) indicates a clear relationship between the room-temperature threshold voltage and the first observed QD electron loading voltage. 
It is well known that transistor threshold voltages increase at cryogenic temperatures, and this is also apparent in the $V_{\rm 1e}$ offset voltage, $\overline{\beta} = \qty{0.21(3)}{\volt}$. 
The intrinsic variation of the first observed electron voltage, which is the expected variation for a fixed $V_{\rm th}$, is $\overline{\sigma} = \qty{16(1)}{\milli\volt}$, which is slightly lower than the observed spread quoted above $\approx$\qty{22}{\milli\volt}.
The excess spread can be attributed to the threshold voltage variation $\overline{V_{\rm th}} = \qty{173(15)}{\milli\volt}$.

\section*{Conclusions}
We have studied a device farm of 1024 quantum dots based on simple transistor structures, but our approach can be extended to more complex unit cells, such as coupled quantum dot systems – the basic building block of semiconductor-based quantum computers. 
Rf readout techniques demonstrated here can be leveraged to embed, compact dispersive spin qubit read-out with the unit cells of scaled up quantum dot architectures~\cite{veldhorst_silicon_2017, Oakes2023}. 
The opportunity for integrated CMOS electronics with qubits reach well beyond addressing, including ultra-low power electronic modules such as digital-to-analogue converters, low-noise amplifiers and digital controllers. These technologies have been demonstrated in standalone processes~\cite{xue_cmosbased_2021}, but tightly integrating all of these modules with semiconductor qubits while retaining their qualities remains an open challenge, especially given the limited cooling power of cryostats and thermal conductivity of silicon at low temperatures. 
Our observation that cryogenic parameters of silicon quantum dot devices can be predicted from room temperature behaviour has important implications on the time and resources required to monitor process variations and optimise the design and production of future quantum devices. Further development of pre-cryogenic methods and analysis tools could allow wider industry engagement and a substantial cost reduction in the technology development, particularly if further correlations can be extracted when complex unit cells are studied.

\section*{Methods}

\subsection*{DC transport measurement}
With a small bias $\sim\qty{1}{\milli\volt}$, the transport current formed by the tunnelling of electrons one-by-one through a QD is typically small $\sim \qtyrange{1}{1000}{\pico\ampere}$.
To record this level of current we use a transimpedance amplifier to convert the current to a voltage with a gain of \qty{e7}{\volt\per\ampere}.
To acquire a Coulomb diamond map, a triangle wave with frequency \qty{20}{\hertz} was applied to the gate and the signal is acquired on the rising edge of the slope.
Devices were measured with either 5 or 10 averages; a single transport Coulomb diamond measurement takes approximately \qty{15}{\second} or \qty{30}{\second}.

\subsection*{RF reflectometry measurement}
Measuring a high-impedance device in reflectometry requires an impedance transforming circuit (see supplementary for more information). 
This allows changes in an otherwise large device impedance to be measured by a \qty{50}{\ohm} matched meter, e.g.~resistance changing from \qtyrange{1000}{100}{\kilo\ohm}.
The complex impedance of the QD has contributions from the resistance and capacitance of the tunnel barrier between the source ohmic and the dot.
When the dot and source electrochemical potentials are equal, the Coulomb blockade is lifted and electrons can tunnel elastically through the barrier. 
This results in an apparent change in the barrier resistance and capacitance, the latter of which may have both quantum and junction contributions~\cite{Luryi1988,Buttiker1993}. 
This impedance change can be detected at the \qty{50}{\ohm} output as a change in the reflected rf signal near the resonance frequency.

\subsection*{Coulomb diamond classification using a convolutional neural network}

We categorise devices into three groups based on their Coulomb blockade maps as shown in Fig.~\ref{fig3}a.
\begin{itemize}
    \item \textit{Good}: 
    These devices display a clearly defined first hourglass in their Coulomb blockade map, enabling extraction of the dot's parameters (see Fig.~\ref{fig1}c).

    \item \textit{Bad}: 
    Devices in this group either lack observable Coulomb blockade or exhibit classical transistor turn-on superimposed with Coulomb blockade. This behavior is likely caused by low resistances of the tunneling barriers.
    
    \item \textit{Multi}: 
    Devices where several series quantum dots are formed can be identified by the overlap of Coulomb diamonds in the Coulomb blockade map, giving rise to extended regions of blockade.
\end{itemize}

The initial classification into three groups was carried out manually by domain experts. 
While this manual approach worked well for the current number of devices, the expected increase in device volume necessitates an automated solution. 
To address this, we trained a convolutional neural network using the \textit{Resnet26d} architecture as a backbone, a standard choice for image classification. 
This architecture comes with weights pre-trained on ImageNet, a vast image database containing millions of images. The use of pre-trained weights is a technique known as transfer learning and allows the network to perform well when trained with our relatively small dataset~\cite{transfer-learning-bengio}.

The Coulomb blockade data is processed by the neural network as grayscale images. 
The dataset was randomly partitioned, with 80$\%$  allocated to a training set and 20$\%$ to a test set. 
During training, the learning rate was scheduled to follow the 1cycle policy introduced by~\cite{smith2019super}. 
Initially, the learning rate increases to a maximum and then gradually decreases. 
Employing a variable learning rate helps regulate the training process, accelerates convergence of the neural network and reduces the network's sensitivity to the learning rate hyperparameter.

To further enhance the neural network's performance, we employ data augmentation techniques like image rotation, warping, zooming, and changes in the saturation.
These techniques introduce variety into the training data, making the model more robust and adaptable to different conditions. 
In addition to data augmentation, we incorporate mixup, a method that generates new images by linearly combining pairs of original training set images~\cite{zhang_mixup_2017}. 
The labels of these new images are determined by the same linear combination of the original labels. 

Furthermore, to minimise over-reliance on training labels, we utilise label smoothing cross-entropy loss. 
In standard cross-entropy loss, training labels are binary, assigning a probability of 1 to the correct class and 0 to others. 
However, with label smoothing, we modify the ground-truth labels by making the correct class probability smaller than 1 and distributing the remaining probability mass uniformly across other classes, helping to reduce model over-confidence~\cite{label-smoothing-hinton}.

The performance of the CNN model is evaluated using a confusion matrix, as shown in Table~\ref{tab:confusion_matrix}, yielding an 88$\%$ accuracy.  
To optimize the neural network further, we suggest a non-binary scoring system for each Coulomb map. 
Here, the network will be trained on labels representing the average consensus from multiple human classifiers, mitigating individual biases.
Additionally, this method will enable the neural network to grasp the intricacies of edge cases between `good' dots and `bad' dots.

\subsection*{Parameter extraction from Coulomb diamonds}
Most key parameters that describe a single QD can be obtained from the position and shape of Coulomb diamond measurements~\cite{hanson_spins_2007, yang_quantum_2020}. 
When measuring the rf response at the drain, the positive (negative) edges of Coulomb diamonds appear when the QD electrochemical potential $\mu_{\rm dot}$ is aligned with the drain (source) electrochemical potential $\mu_{\rm D}$~$(\mu_{\rm S})$. 
The first observed electron loading voltage $V_{\rm 1e}$ corresponds to the crossing point of the first pair of these edges (nominally at $\mu_{\rm S}=\mu_{\rm D}=\qty{0}{\electronvolt}$).

The gate lever arm $\alpha_\mathrm{G} = C_\mathrm{G} / C_\Sigma$, where $C_\mathrm{G}$ is the gate capacitance and $C_\Sigma$ is the sum of the dot capacitance to each terminal, represents the coupling strength of the gate to the dot. 
This parameter can be calculated as 
\begin{equation}
    \alpha_{\rm G} = \left(\frac{1}{\abs{m_1}} + \frac{1}{\abs{m_2}}\right)^{-1},
\end{equation} with $m_1$ and $m_2$ being the positive and negative gradients of the pair of edges that form the first hourglass (see Fig.~\ref{fig1}c).

Moreover, when a voltage bias is applied antisymmetrically ($V_\mathrm{D} = -V_\mathrm{S}$), the relative coupling capacitance of the source $\alpha_\mathrm{S} = C_{\rm S}/C_\Sigma$ and drain $\alpha_\mathrm{D}  = C_{\rm D}/C_\Sigma$ can be directly obtained from  the gradient of the Coulomb diamond edges as 
\begin{equation}
    \alpha_{\rm D} - \alpha_{\rm S} = \frac{m_1 + m_2}{m_1 - m_2}.
\end{equation}
This is a measure of the asymmetry of dot formation under the gate.

The full step-by-step process for extracting dot parameters from Coulomb blockade maps is detailed in the supplementary, but we provide a succinct summary here.
We perform digital filtering to reduce the noise and enhance contrast in the acquired data, and then apply a Canny edge detection algorithm to digitise the charge stability map and identify the edges of the Coulomb diamonds.
We then employ a Hough transform to convert the binary image to information about the edges parametrised by their length and angle.
We identify good fits to the first Coulomb oscillation with a pair of long line segments, one with a positive slope and the other negative, that intersect near to $V_{\rm DS} = \qty{0}{\volt}$.

\subsection*{Room temperature correlations to cryogenic behaviour}
To understand the systematic relationship between $V_{\rm 1e}$ and $V_{\rm th}$ and their random variation, we employ the probabilistic programming framework numpyro~\cite{phan2019composable}.
As stated in the main text, we use a model defined by three random variables: $\alpha$, $\beta$ and $\sigma$.
These variables are initially set with a suitable prior distribution.
The posterior distribution is then formed using Hamiltonian Monte Carlo~\cite{Hoffman2011} sampling.

We also consider an alternative model motivated by the distinctive asymmetry in the distribution in Fig.~\ref{fig:room-temperature-correlation}b.
This second model treats $V_{\rm 1e}$ as a mixture of two distinct normal distributions. 
The two means are determined by a different linear dependence on $V_{\rm th}$, with shared slope $\alpha$ but different intercept $\beta$.
The weighting of these distributions in the mixture is also treated as a random variable.

To evaluate the performance of each model we use the leave one out cross validation technique (LOO-CV)~\cite{Vehtari2015}. 
This measures the performance of the model by taking average log-likelihood for the observed data across all samples from the posterior distribution. 
In addition, we measure the Pareto-$k$ diagnostic for all observed data to verify that the measured LOO-CV values are reliable. 
On the observed data, we note that there is no significant difference between the models as determined by their LOO-CV score, with both results returning values of between $-485$ and $-490$ and expected error of $\approx7$ dependent on prior selection. 
This suggests that with the current data there is no reason to select one model over the other, and so we present the simpler linear model. 

\subsection*{Device layout}
Best layout practices~\cite{Sharma2021} were followed to minimize the impact of well-known sources of manufacturing variability~\cite{Ndiaye2016}. The placement of each DUT within the 32 $\times$ 32 array is chosen to minimize layout dependent effects such as length of oxide diffusion (LOD), well-proximity effect (WPE) and oxide spacing effect (OSE). 
Additionally we avoid using lower level metal layers for routing over or near the gates of all the DUTs. 
Furthermore, we use a placement algorithm for each set of instances (devices with the same dimensions) to minimize the impact of 2-dimensional patterns across the farm in the extracted parameters. 
The algorithm combines randomisation of the placement of each instance within the array, with the requirement for each set to have a common centroid located at the centre of the array. 
The resulting device placement can be seen in Ext.~Fig.~\ref{fig:farm-map}.

\subsection*{Device fabrication}
The die was fabricated using the GlobalFoundries 22FDX \qty{22}{\nano\metre} FD-SOI process.

\subsection*{Experimental setup}
The integrated circuit die was glued to a carrier printed-circuit board (PCB) with conducting silver paste. 
The die is wire-bonded (\qty{17.5}{\micro\metre} AlSi) to gold-plated copper tracks. 
The low-frequency control lines are routed to a connector to attach to a motherboard which has passive first-order in-line filtering ($RC = \qty{10}{\micro\second}$).
The reflectometry line is directly routed to an SMP connector on the carrier PCB.

All cryogenic measurements were performed in a Bluefors XLD dilution refrigerator, where the device was mounted to the mixing chamber plate operating at \qty{10}{\milli\kelvin}.
When the chip is powered on, the die temperature was measured as \qty{600}{\milli\kelvin} using on-chip thermometry~\cite{noah_cmos_2023} (value quoted for a nominally identical die).

A QDevil QDAC II was used to supply the DC voltages to the QD device terminals, chip supplies and row, column address lines. 
To sweep the gate voltage, a triangle wave with 50\% duty cycle was supplied by a Keysight 33500B arbitrary waveform generator. 
The radio-frequency reference signal was generated by a Rohde \& Schwarz SMB100B.
The reflected signal was amplified by a Low Noise Factory cryogenic amplifier (LNF-LNC0.2\_3A s/n 2541z) at \qty{4}{\kelvin}.
This signal is then further amplified at room temperature using two Mini-circuits amplifiers (ZX60-P103LN+ and ZX60-33LNR-S+) before being separated into its in-phase and quadrature components using a Polyphase microwave quadrature demodulator (AD0540B).
These signals are finally amplified and filtered by a Stanford SR560 and then digitized via a Spectrum M4i.4421-x8 digitizer PCIe card.

For dc measurements we used two transimpedance amplifiers (Basel Precision Instruments SP983c, IF3602 JFET) to simultaneously monitor the source $I_{\rm S}$ and drain $I_{\rm D}$ currents.
The gain of the amplifiers was set to \qty{e7}{\volt\per\ampere} with a low-pass filter bandwidth of \qty{1}{\kilo\hertz}.
For our measurements we use $I_{\rm sig} = (I_{\rm D} - I_{\rm S})/2$ to remove any offsets.
During rf measurements, the transimpedance amplifiers are removed and the source and drain are directly driven with the QDAC II voltage sources above.

\section*{Data availability}
The data that support the plots within this paper and other findings of this study are available from the corresponding authors upon request.

\section*{Acknowledgements}
E.J.T.\ acknowledges the Engineering and Physical Sciences Research Council (EPSRC) through the Centre for Doctoral Training in Delivering Quantum Technologies [EP/S021582/1]. M.F.G.Z.\ acknowledges a UKRI Future Leaders Fellowship [MR/V023284/1].
The authors thank Nigel Cave at GlobalFoundries for helpful discussions.

\section*{Author contributions}
E.J.T., V.N.C.T.\ and D.P.\ performed the experiments under the supervision of M.A.I.J.\ and M.F.G.Z.;
M.d.K.\ and D.J.I.\ contributed to the preparation of the experiment and preliminary characterisation;
M.F.G.Z.\ and J.J.L.M.\ conceived the experiment;
A.G.S.\ designed the device with contributions from M.F.G.Z.\ and J.J.L.M.; the device analogue components were validated by G.M.N.;
D.F.W.\ and V.N.C.T.\ designed and developed the automated analysis tools;
Further data analysis was performed by E.J.T., V.N.C.T., D.F.W.\ and M.A.I.J.;
The manuscript was prepared by M.A.I.J.\ with contributions from all authors.

\section*{Competing interests}
The authors declare no competing interests.

\section*{Extended Data}
\begin{table}[htbp]
    \centering
    \caption{Confusion matrix of the convolutional neural network on the test data set. The elements on the diagonal are the number of instances correctly identified as `good' or `other`, whereas the off-diagonal elements are the false `good' and `other'. We observe a precision of 0.67 and a recall of 0.86.}
    \label{tab:confusion_matrix}
    \renewcommand{\arraystretch}{1.5}
    \begin{tabular}{cc|cc}
    \multicolumn{2}{c}{}
            &   \multicolumn{2}{c}{\textbf{CNN}} \\
    &       &   Good &   Other              \\ 
    \cline{2-4}
    \multirow{2}{*}{\rotatebox[origin=c]{90}{\textbf{Manual}}}
    & Good   & 37   & 6                 \\
    & Other    & 18    & 143                \\ 
    \cline{2-4}
    \end{tabular}
    
\end{table}

\begin{figure}[t]%
    \centering
    \includegraphics[width=\linewidth]{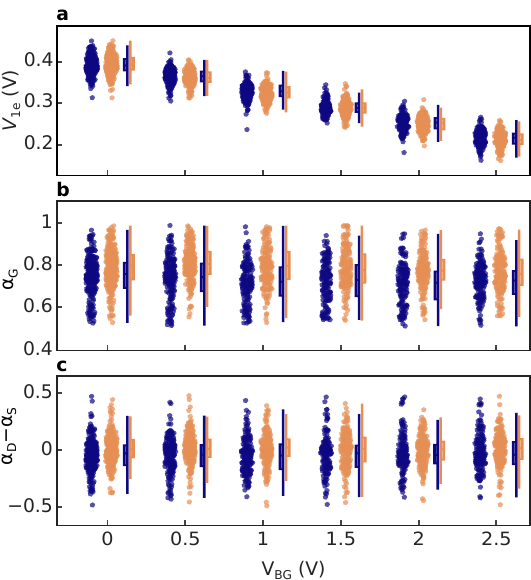}
    \caption{
        \textbf{a-c}, Estimated distributions of first observed electron loading voltage, gate lever arm and QD asymmetry as back-gate voltage is varied ($L = \qty{28}{\nano\metre})$. 
        A swarm plot is shown alongside a box plot to provide key information about the data.
        Each point in the swarm plot is a single observation, and each point is laterally spaced to minimise overlap with others.
        The blue data is drawn from rf measurements, while the orange data is drawn from dc measurements. \textbf{a} shows a reduction in $V_{\rm 1e}$ with increasing back-gate voltage, caused by the shifting of the conduction band edge toward the Fermi level. 
        \textbf{b} shows a small reduction in average lever arm with increasing back-gate, matching the expectation that the electron moves further from the top-gate. 
        \textbf{c} shows a minimal effect from back-gate voltage on $\alpha_\mathrm{D} - \alpha_\mathrm{S}$, with only a small increase in standard deviation notable. This indicates that the QDs remains centred between the leads. 
    }
    \label{fig:back-gate}
\end{figure}

\begin{figure}[t]
    \centering
    \includegraphics{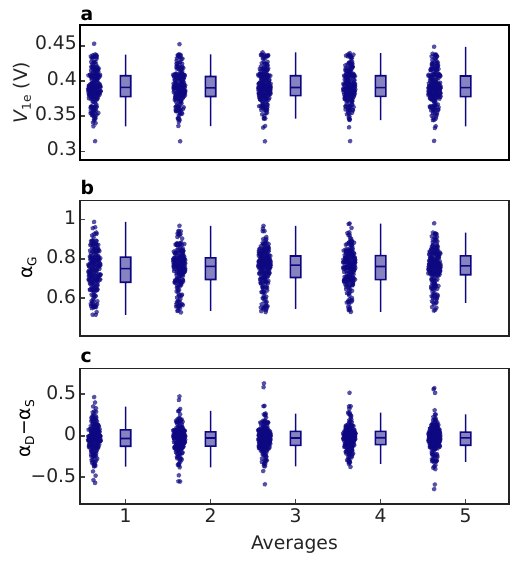}
    \caption{
        \textbf{a-c}, Distributions of extracted parameters (first detected electron loading voltage, gate lever arm, and source-drain asymmetry) using swarm plots (left) and a box plot (right), as the number of averages is varied in rf measurements. The swarm plot reveals individual data points' distribution, while the box plot displays central tendencies and quartiles for easy comparison and outlier detection. As expected, we observe a decrease in the parameter standard deviation as averages increase. Remarkably, parameters are extracted accurately even with a single average, giving the results in Fig.~\ref{fig3} in the main text.
    }
    \label{fig:averages}
\end{figure}

\begin{figure*}
    \centering
    \includegraphics[width=\linewidth]{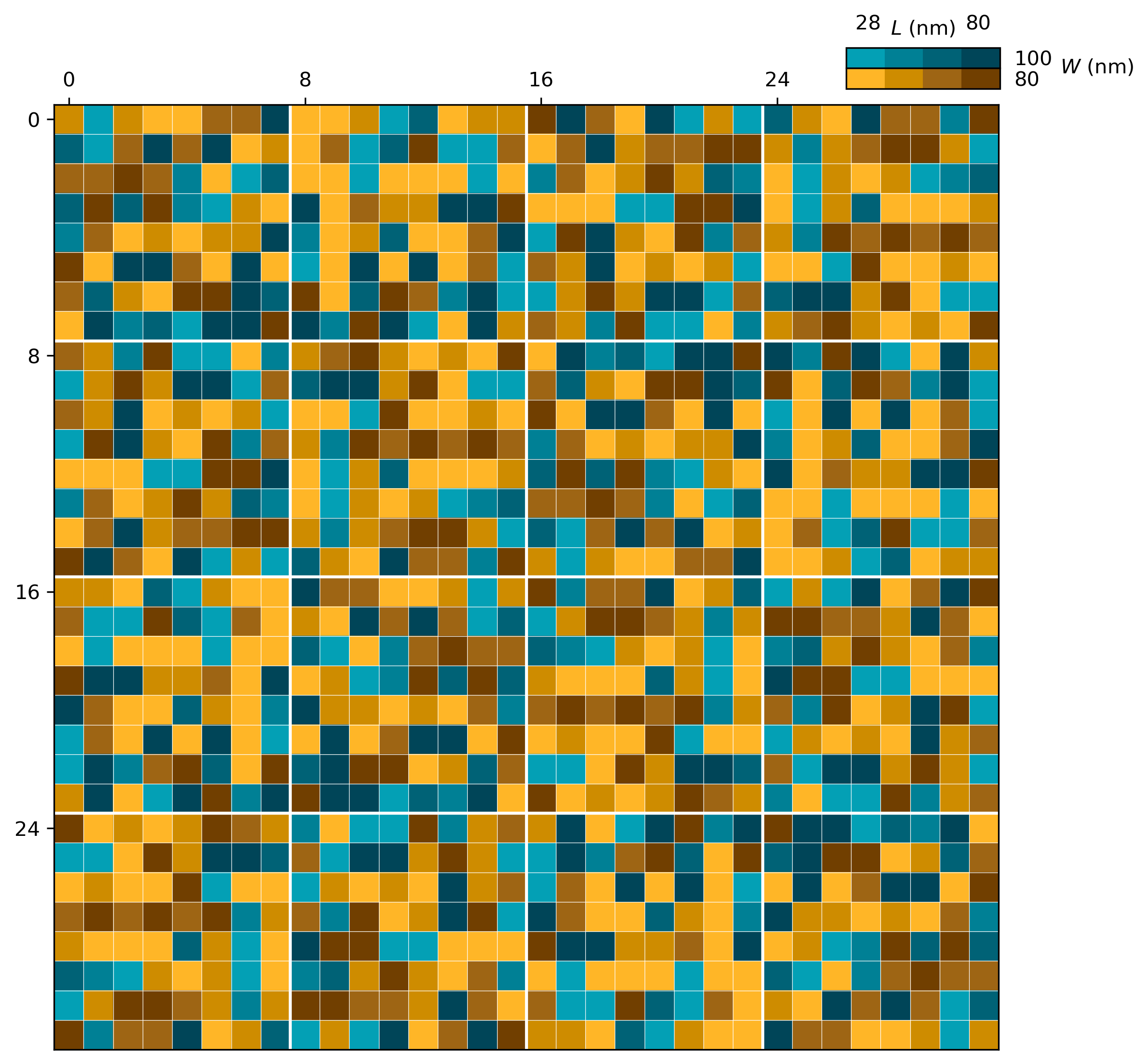}
    \caption{The layout position of devices in the farm.
    All identical instances form a set, and each instance within a set is placed using randomisation with the centre of the array as the centroid.
    }
    \label{fig:farm-map}
\end{figure*}

\clearpage
\bibliography{main}

\end{document}


\title{Rapid cryogenic characterisation of 1024 integrated silicon quantum dots\\Supplementary materials}

\author{Edward J.~Thomas}
\affilqmt
\affilucl

\author{Virginia N.~Ciriano-Tejel}
\affilqmt

\author{David F.~Wise}
\affilqmt

\author{Domenic Prete}
\affilqmt

\author{Mathieu de Kruijf}
\affilqmt
\affillcn

\author{David J.~Ibberson}
\affilqmt

\author{Grayson M.~Noah}
\affilqmt

\author{Alberto Gomez-Saiz}
\affilqmt

\author{M.~Fernando Gonzalez-Zalba}
\affilqmt

\author{Mark A.~I.~Johnson}
\affilqmt

\author{John J.~L.~Morton}
\affilqmt
\affilucl
\affillcn

\maketitle

\section{Experimental design}
\subsection{Apparatus}
\begin{figure}[p]
    \centering
    \includegraphics[width=0.9\textwidth]{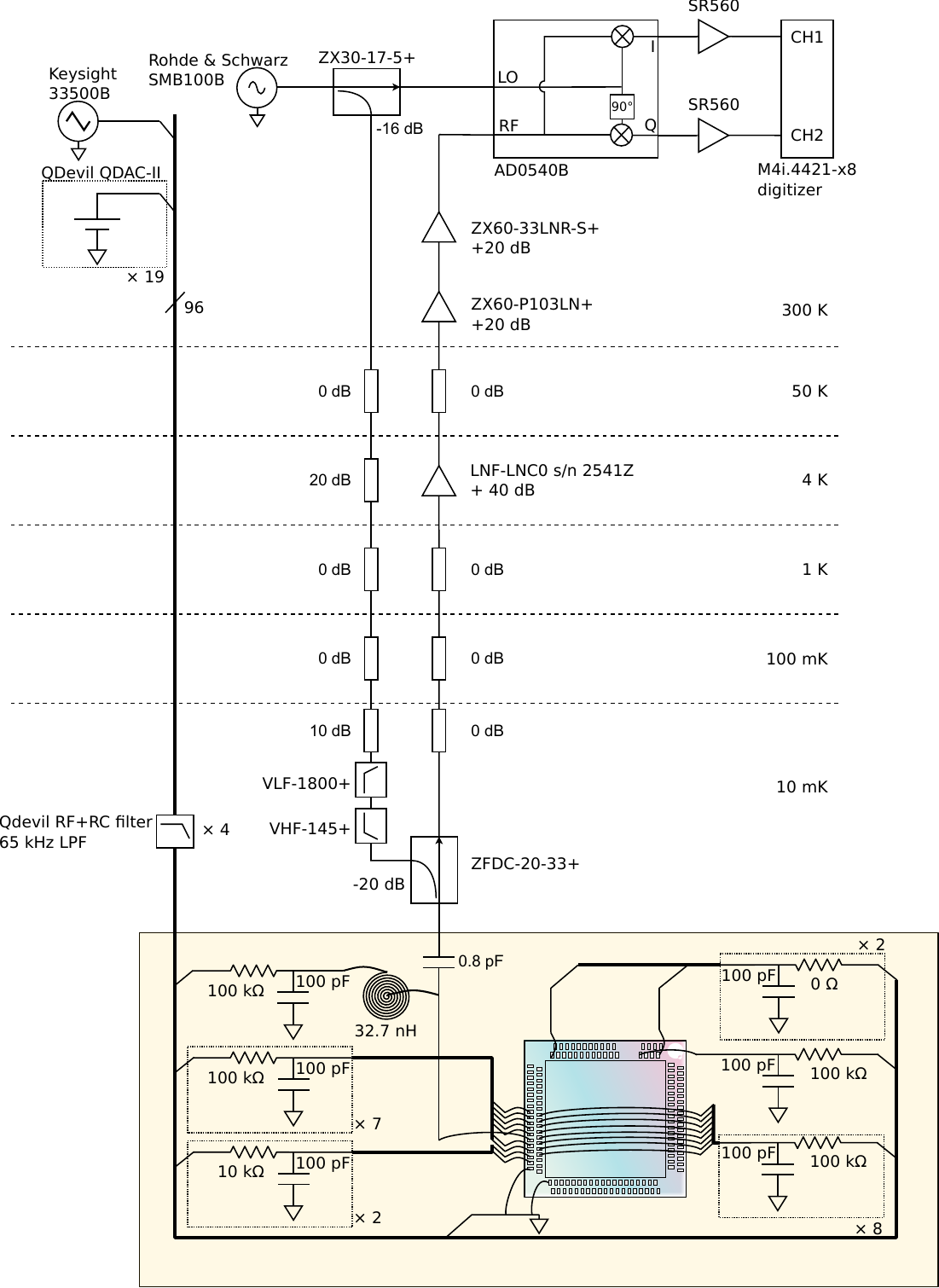}
    \caption{
    The experimental apparatus used for rf reflectometry measurements includes room temperature electronics which are connected to the device at the mixing chamber plate via twisted-pair looms for low frequency signalling, and to a coaxial line for rf reflectometry.
    The device is mounted on a PCB (yellow region) with on-board RC filtering; the device is connected to the signal lines via wire bonds.}
    \label{fig:fridge_setup}
\end{figure}

The experimental apparatus is shown in Fig.~\ref{fig:fridge_setup}.
All dc biasing is provided via a set of 96 twisted-pair phosphor bronze loom wires, which are filtered with QDevil RC+RF filters. 
These lines are then filtered on-board using \qty{100}{\kilo\ohm} series resistance and \qty{100}{\pico\farad} shunt capacitance except for the multiplexer n- and p-wells, used for back-biasing the transmission gates, which have \qty{10}{\kilo\ohm} series resistance and the die power supply lines with \qty{0}{\ohm} series resistance.

The rf lines between temperature stages are CuNi coaxial cables with fixed attenuators for thermalisation and noise reduction.

\subsection{Multiplexer design and validation}
\begin{figure}
    \centering
    \includegraphics[width=0.6\textwidth]{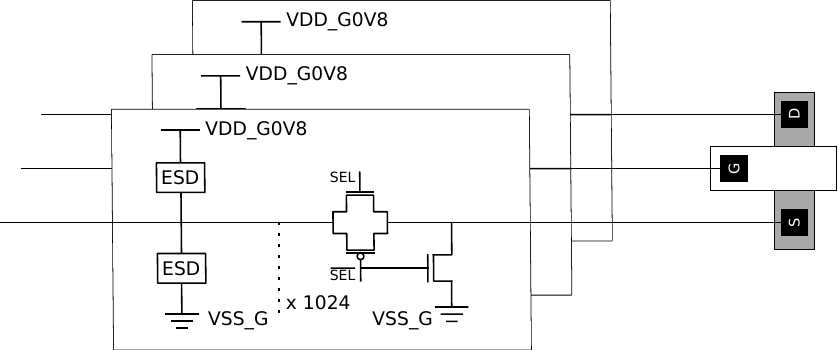}
    \caption{Schematic representation of bond pad cell with ESD protection and the multiplexer structure in die.
    Here three separate device contacts are accessed through nominally identical multiplexing structures via bond pads.
    }
\label{fig:mux_schematic}
\end{figure}

Fig.~\ref{fig:mux_schematic} shows the structure of the multiplexer used in this report.
The bond pad associated with a device contact (e.g. the gate) sees an ESD protection circuit and the input of 1024 transmission gates (back-to-back NFET and PFETs).
When the device is powered up, one transmission gate will be open and all others closed, determined by the chosen row and column address. 
The device select signals are generated by a digital one-hot decoder.
For all deselected devices a pull-down transistor is active ensuring these devices are tied to the global ground (VSS\_G).

\subsection{Resonator design and validation}
\begin{figure}
    \centering
    \includegraphics[width=0.5\textwidth]{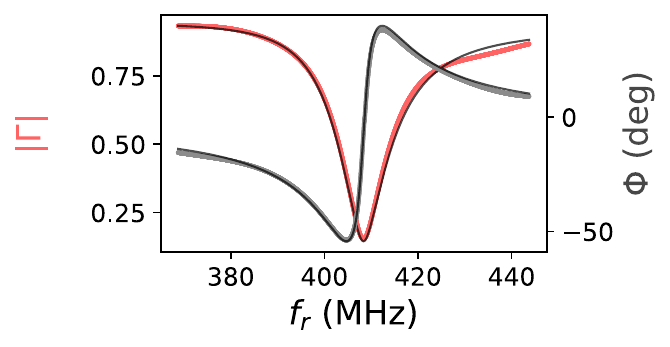}
    \caption{Reflection coefficient of the multiplexer chip bonded to the resonator (magnitude $|\Gamma|$ (red) and phase $\Phi$ (grey)) at 1K. We observe a natural resonant frequency of $f_{\rm r}=\qty{407}{\mega\hertz}$).}
    \label{fig:resonator}
\end{figure}
As shown in Fig.~\ref{fig:fridge_setup}, we have adopted the conventional approach~\cite{petersson_charge_2010, schroer_radio_2012, schaal_cmos_2019, gonzalez-zalba_probing_2015} of embedding the device in a lumped element LC resonant circuit. 
In this setup, we have an NbTiN superconducting spiral inductor comprising 8 turns, fabricated by StarCryo, with an inductance value of $L = \qty{32.7}{\nano\henry}$. 
This inductor is connected in parallel with the device channel. 
Additionally, a coupling capacitor $C_\mathrm{C} = \qty{0.8}{\pico\farad}$ is responsible for coupling the external circuitry to the device and resonator. 
The total capacitance of the resonator, $C_{\rm total} = \qty{4.66}{\pico\farad}$ is determined from its natural frequency $f_{\rm r} = \qty{407.2}{\mega\hertz}$. 
This capacitance is equal to $C_{\rm total} = C_{\rm C} + C_{\rm p,\,chip} + C_{\rm p,\,PCB}$, where $C_{\rm p,\,chip}$ and $C_{\rm p,\,PCB}$ are the parasitic capacitance of the chip (including the multiplexer and the ESD structure) and the parasitic capacitance of the PCB, respectively.
By comparing the resonant frequency of the resonator before and after bonding the multiplexer chip, we have determined that the chip's parasitic capacitance is $C_{\rm p,chip} = \qty{0.8}{\pico\farad}$.

Figure~\ref{fig:resonator} depicts the resonator's response upon bonding to the chip at 1K. We observed a matching coefficient $\beta = 0.66$, a total quality factor of $Q_{\rm L} = 21$, an internal quality factor $Q_{\rm i} = 34.9$, and a resonator impedance $Z_{\rm r} = \qty{75} \Omega$ (determined following~\cite{probst2015}). Utilizing this resonator, we achieve a minimum integration time of $t_{\rm min} = \qty{0.16}{\nano\second}$, by extrapolating the linear dependence of the SNR with integration time down to SNR$=1$. 
We note that the bandwidth of the resonator gives a rise time $\approx \qty{20}{\nano\second}$, and although techniques could be used to reduced this minimum response time by pulsed readout~\cite{walter_rapid_2017} it is ultimately likely to limit the minimum readout time. 
Having said that, given that the observable is conductance, the resonator can be designed to have higher bandwidth without deteriorating the SNR~\cite{schoelkopf_radiofrequency_1998}.

\clearpage

\section{Data analysis pipeline}
\begin{figure}
    \centering
    \includegraphics[width=\columnwidth]{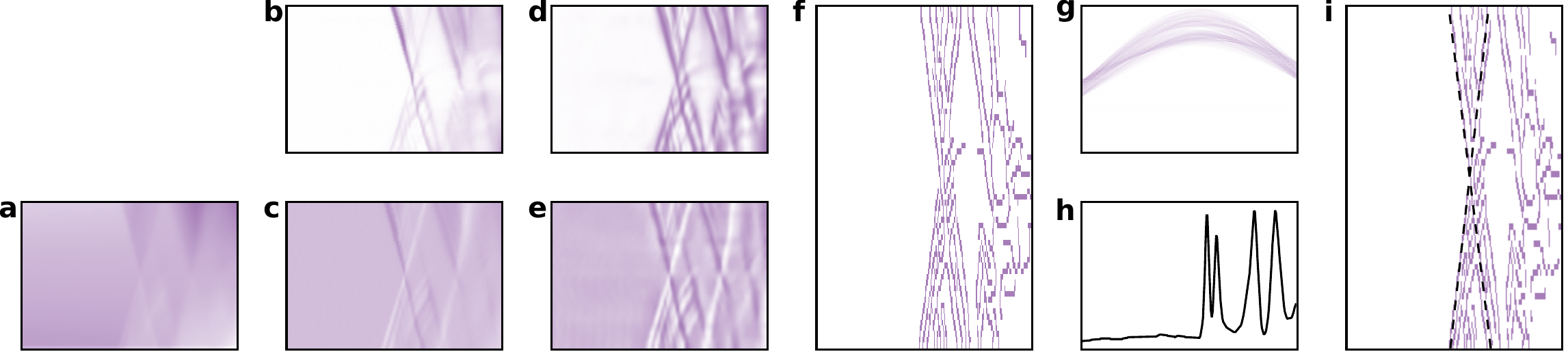}
    \caption{Image processing pipeline for automated parameter extraction.}
    \label{fig:pipeline-fig}
\end{figure}

Data analysis is a multi-step process that combines machine learning techniques for measurement classification with traditional image processing for parameter extraction, an example of the procedure is shown in Fig.~\ref{fig:pipeline-fig}. 
To allow for a single analysis pipeline for both dc and rf data, dc measurements (shown in Fig.~\ref{fig:pipeline-fig}a) are first numerically differentiated to give $\partial  I_{\rm D}/\partial V_{\rm DS}$ which can be analysed identically to rf measurements. 

Following this, the first step is to use a trained neural network to identify whether a measurement shows identifiable Coulomb blockade from which parameters can be extracted, as discussed in the main text. 
Once identified as suitable, the map goes through a simple image processing pipeline:
\begin{enumerate}
    \item Background drift is removed row-by-row using the mean value of the first 100 values of each row, shown in Fig.~\ref{fig:pipeline-fig}b and c for rf and dc data, respectively.
    \item The image contrast is enhanced using contrast limited adaptive histogram equalisation (CLAHE), shown in Fig.~\ref{fig:pipeline-fig}d-e for rf and dc data, respectively.
    \item Canny edge detection~\cite{canny1986computational} is used to detect the edges of Coulomb diamonds and binarise the image (all pixel values either 1 or 0), shown in Fig.~\ref{fig:pipeline-fig}f.
    \item The Hough transform~\cite{duda1972use} is used to extract all possible straight lines in the image that exceed a certain length threshold, an example of its output is shown in Fig.~\ref{fig:pipeline-fig}g.
    \item Coulomb oscillations at $V_{\rm DS} = \qty{0}{\volt}$ are detected using a peak finding algorithm and their $V_{\rm G}$ values recorded, shown in Fig.~\ref{fig:pipeline-fig}h.
    \item Lines are divided into those with positive and negative gradients and all possible negative/positive pairs are scored according to:
    \begin{enumerate}
        \item Total line length
        \item Proximity of their crossing point to $V_{\rm DS} = \qty{0}{\volt}$
        \item Proximity of the crossing point to a Coulomb oscillation in $V_{\rm G}$
        \item Similarity of the absolute value of their gradients
        \item How low in gate voltage their crossing point is (to prioritise the first visible oscillation)
    \end{enumerate}
    \item Line pairs that show non-physical parameters (e.g. lever arm > 1) are filtered out. We also discard pairs with lever arm < 0.5, as these are deemed highly unlikely in this architecture 
    \item Scores are standardised and weighted and the line pairs are ranked according to the sum of scores, with the highest scoring pair used to extract dot parameters as shown in Fig.~\ref{fig:pipeline-fig}i
\end{enumerate}

\section{Assessing layout dependent effects via quantum dot quality}
As stated in the main text, during the design and layout of the farm devices known sources of variation in the semiconductor manufacturing process were taken into account to minimise any bias in the device statistics.
To minimise layout-dependent effects, identical device designs were placed in the farm using a semi-random algorithm.
We find that the device design is a stronger indicator of its measured quality than its location within the farm, as evidenced by the nearly uniform histograms of quality as organised by row or column.
We calculate the mean Kullback-Liebler divergence (KLD) of these histograms from a uniform distribution to be \num{0.031+-0.017}, while the mean KLD of the histograms organised by device type is \num{0.45}.
\begin{figure}[htbp]
    \centering
    \includegraphics[]{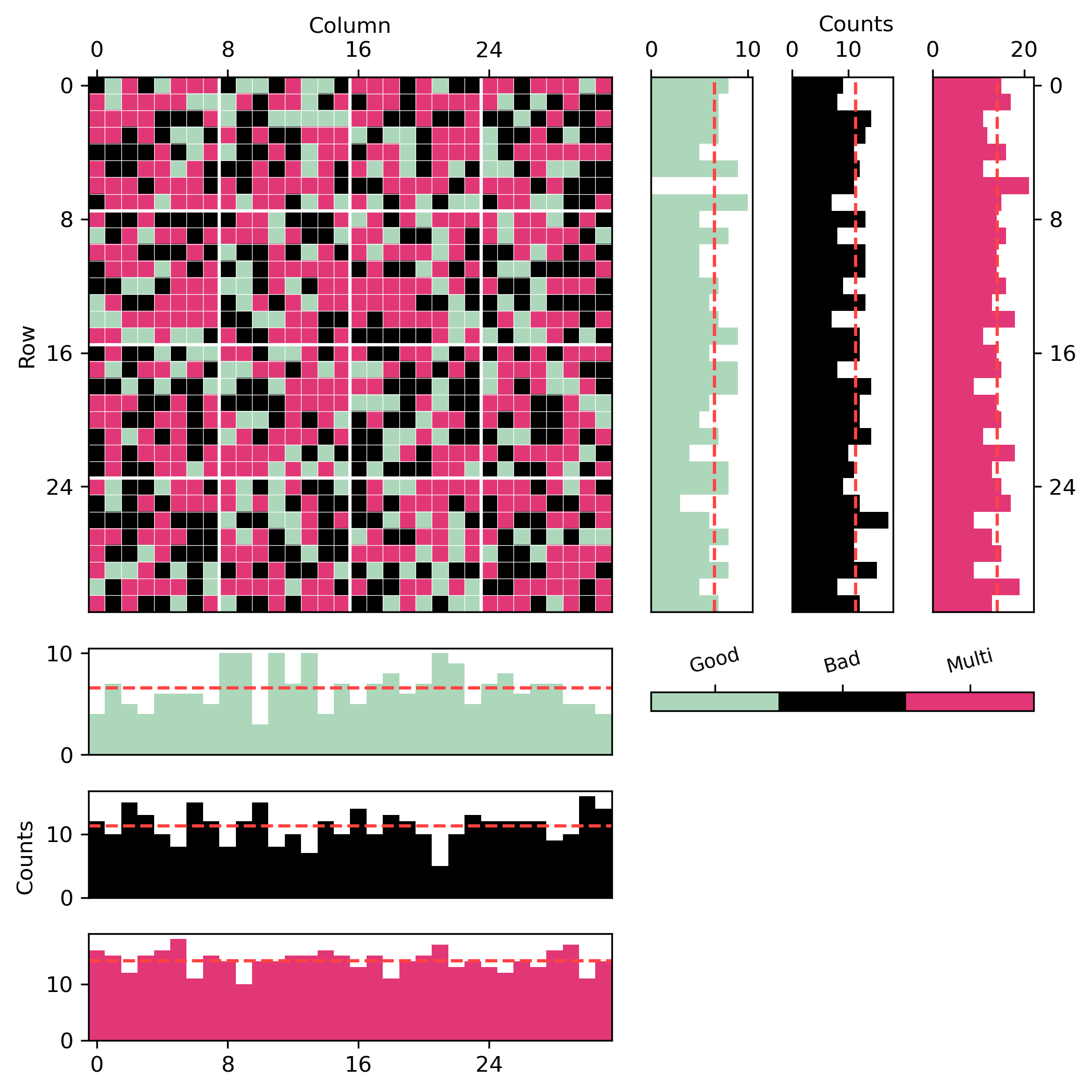}
    \caption{A map of all devices in the farm and their neural network classification, with corresponding row and column histograms for each classification.
    The red dashed lines indicates the probability density of a uniform distribution with the same weight as each histogram.
    }
    \label{fig:quality_heat_map}
\end{figure}
\clearpage
\section{QD lever arm - asymmetry relationship}
\begin{figure}[htbp]
    \centering
    \includegraphics{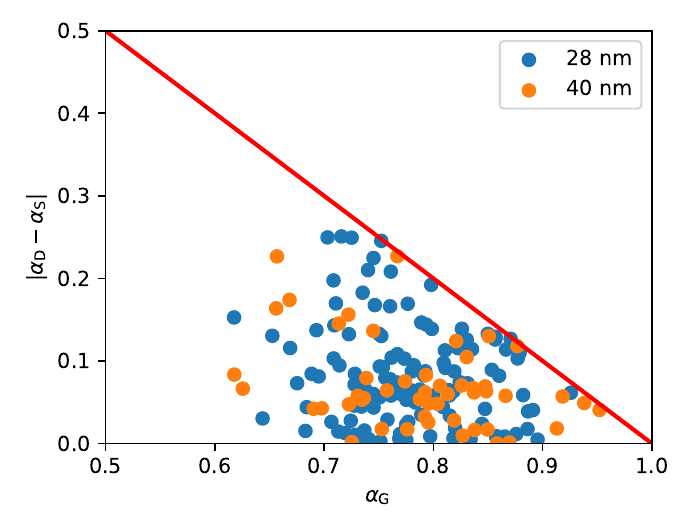}
    \caption{A correlation map showing the bounded relationship between the QD asymmetry and the gate lever arm. 
    The red line is a physical upper bound given by the unity sum of the lever arms.
    There is a predictive, causal relationship between a large measured gate lever necessarily having a small QD asymmetry in absolute terms.
    }
    \label{fig:lever_arm_asymmetry}
\end{figure}

For an SET with 3 device contacts, source (S), drain (D) and gate (G), the lever arm of each contact to the QD must sum to unity,
\begin{equation}
    \alpha_{\rm S} + \alpha_{\rm D} + \alpha_{\rm G} = 1.\label{eq:lever_arm_sum}
\end{equation}
In the main text two parameters we have extracted from our Coulomb diamonds are $\alpha_{\rm G}$ and $\alpha_{\rm D} - \alpha_{\rm S}$.
One important aspect of any automated fitting process is that it returns physical results, so we now consider how Eq.~\ref{eq:lever_arm_sum} constrains these two parameters.
Our results reveal (Fig.~\ref{fig:lever_arm_asymmetry}) that a large absolute asymmetry $\abs{\alpha_{\rm D} - \alpha_{\rm S}}$ implies that the gate lever arm must be less markedly less than 1.

This can be understood by considering how the difference of lever arms relates to the sum of the lever arms.
Since the lever arms are positive, it follows that $\abs{\alpha_{\rm D} - \alpha_{\rm S}} \le \alpha_{\rm S} + \alpha_{\rm D}$.
By substituting into Eq.~\ref{eq:lever_arm_sum} we can define a physical constraint (which is indicated in Fig.~\ref{fig:lever_arm_asymmetry} by the red line) on our extracted parameters $\alpha_{\rm D} - \alpha_{\rm S}$ and $\alpha_{\rm G}$,
\begin{gather}
    \abs{\alpha_{\rm D} - \alpha_{\rm S}} + \alpha_{\rm G} \le 1, \\
    \alpha_{\rm G} - 1 \le \alpha_{\rm D} - \alpha_{\rm S} \le 1 - \alpha_{\rm G}.
\end{gather}

\section{Room temperature measurements}

We analyse the transport measurements through the transistors at room temperature with a drain-source bias of $V_{\rm DS} = \qty{50}{\milli\volt}$. 
The threshold voltage ($V_{\rm th}$) is determined by extrapolating the $I_{\rm D}$-$V_{\rm GS}$ curve, between the maximum transconductance point $\max(g_m) = \max{(d I_{\rm D} / d V_{\rm GS})}$ and the maximum subthreshold slope point $\max(SS) = \max{(\log(I_{\rm D}) /  V_{\rm GS})}$~\cite{Pierre2010}. 
$V_{\rm th}$ is the voltage where this extrapolated line intersects $I_{\rm D} = 0$.

\section{Assessment of tunnel rates}
The tunnel rate can be extracted from a resonant tunnelling measurement by observing the Lorentzian broadening of the peak.
The reflectometry measurements shown here were performed with a large probe power ($\sim$\qty{-65}{\deci\bel\of{m}}) to ensure a high signal to noise ratio was achieved.
This results in the resonant tunneling features appearing broadened in the gate voltage axis, which precludes an assessment of the tunnel rate directly since the power broadening obscures the underlying Lorentzian broadening.

Alternatively, the dc current $I_{\rm D}$ formed through sequential tunnelling of electrons from the source to the drain via the dot can be described by a gross tunnelling rate $\Gamma$.
This rate is the average number of electrons that flow from the source to the drain, and is the harmonic mean of the individual tunnel rates from the dot to the source ($\Gamma_{\rm S}$) and drain ($\Gamma_{\rm D}$) 
\begin{equation}
    \Gamma^{-1} = \Gamma_{\rm D}^{-1} + \Gamma_{\rm S}^{-1}.
\end{equation}
Since the gross tunnel rate can be directly measured in our devices in dc transport $\Gamma = I_{\rm D}/{\abs{e}}$, we can use this to define a lower-bound for the individual tunnel rates.
From our measurements, we extract the dc drain current $I_{\rm D}$ under the bias conditions $V_{\rm GS} = V_{\rm 1e}$ and $V_{\rm DS} = \qty{1}{\milli\volt}$, giving $I_{\rm D} =  \qty{1.2+-0.9}{\nano\ampere}$ across all devices.
This gives a mean gross tunnel rate of $\Gamma = \qty{7.4e9}{\per\second}$.

\section{Outlook on device yield}
We now address the imperfect yield of good QD features in the devices tested here.
Good QD features require two main ingredients:
\begin{itemize}
    \item A potential well that can confine at least one electron
    \item Potential barriers between the potential well and the device leads of sufficient length and height to ensure the tunnelling resistance is greater than the resistance quantum $R_Q = h / e^2$.
\end{itemize}
Specific to our measurements using rf reflectometry, we also require that the tunnel rate is larger than the probe frequency to observe a signal, i.e.~$\Gamma/2\pi > \qty{407}{\mega\hertz}$, which corresponds to a current of $\sim\qty{60}{\pico\ampere}$.

For devices where two ohmic leads are gated by a single electrode, as in a typical field-effect transistor (FET), charge confinement is only possible in a narrow operating range due to the lack of control over the tunnel barriers between the leads and the formed QD \cite{sellier2007subthreshold,yang2020quantum}.
Outside of this region the device is either pinched-off (no QD formed and no conductance between the leads) or the device exhibits either direct tunnelling or co-tunnelling from one lead to the other, resulting in a background flow of electrons which obscures charge confinement.
With a single electrode it is impossible to independently control QD formation and the tunnel barriers.
To circumvent this, multiple electrodes are typically used which define QD gates and barrier gates, or confinement gates, to enable this independent control.

An alternative approach is to create larger intrinsic sections between the leads and the gated region. 
In a self-aligned lithography process, after depositing the gate electrode, the leads are defined by implanting dopants into the active silicon region.
Dopants are screened by the gate electrode as well as the spacers (see main text Fig.~1), which enables gate control over the channel below.
By controlling the size of the spacers, we can push the doped regions further from the site where the QD will form\cite{wacquez2010single}. 

\section{Charging energy}

We obtain a charge energy of $E_C = \qty{15+-4}{\milli\electronvolt}$ as $E_C = |e|\Delta V_{\rm G}  \alpha_{\rm G}$, where $\abs{e} \approx \qty{1.6e-19}{\coulomb}$ is the elementary charge, $\alpha_{\rm G}$ is the gate lever arm and  $\Delta V_{\rm G}$
is the difference between the first and second detected electron ($V_{\rm 2e}$) loading voltages. 
Some examples of Coulomb diamonds can be found in the main text showing slightly smaller charging energies than the average. 

\bibliography{supp}